\documentclass[aps,prb,twocolumn,showpacs,superscriptaddress]{revtex4}

\usepackage{graphicx} 

\begin{document}

\title
{Confinement of electrons in size modulated silicon nanowires}

\author{S. Cahangirov}
\affiliation{UNAM-Institute of Materials Science and
Nanotechnology, Bilkent University, Ankara 06800, Turkey}
\author{S. Ciraci} \email{ciraci@fen.bilkent.edu.tr}
\affiliation{UNAM-Institute of Materials Science and
Nanotechnology, Bilkent University, Ankara 06800, Turkey}
\affiliation{Department of Physics, Bilkent University, Ankara
06800, Turkey}
\date{\today}

\begin{abstract}

Based on first-principles calculations we showed that
superlattices of periodically repeated junctions of hydrogen
saturated silicon nanowire segments having different lengths and
diameters form multiple quantum well structures. The band gap of
the superlattice is modulated in real space as its diameter does
and results in a band gap in momentum space which is different
from constituent nanowires. Specific electronic states can be
confined in either narrow or wide regions of superlattice. The
type of the band lineup and hence the offsets of valence and
conduction bands depend on the orientation of the superlattice as
well as on the diameters of the constituent segments. Effects of
the SiH vacancy and substitutional impurities on the electronic
and magnetic properties have been investigated by carrying out
spin-polarized calculations. Substitutional impurities with
localized states near band edges can make modulation doping
possible. Stability of the superlattice structure was examined by
ab initio molecular dynamics calculations at high temperatures.

\end{abstract}


\maketitle
\section{Introduction}
Rod-like Si nanowires (SiNW) have been synthesized down
to $\sim$ 1 nm diameter.\cite{Ma} They are attractive
one-dimensional (1D) materials because of the well-known silicon
fabrication technology that make them directly usable on the
Si-based chips. Even if unsaturated dangling bonds on the outer
surface usually attribute a metallic character to SiNWs, they
become insulator (or semiconductor) upon saturation of these
dangling bonds by hydrogen atoms.\cite{Singh} SiNWs display
diversity of electronic properties depending on their diameter, as
well as their orientation. In particular, the band gap of
semiconductor SiNWs varies with their diameters. They can be used
in various electronic, spintronic and optical applications, such
as  field effect transistors \cite{Cui}, light emitting diodes
\cite{Huang}, lasers \cite{Duan} and interconnects. The
conductance of these semiconductor nanowires can be tuned easily
by doping\cite{dopex,Serra} during the fabrication process or by
applying a gate voltage. Recent studies have shown that 3{\it{d}}
transition metal doped Si nanowires become
half-metallic.\cite{halfmetal}

This letter demonstrates that SiNWs of different diameters can
form stable superlattices. The electronic band structure of the
superlattice is different from the constituent SiNWs and is
modulated in real space leading to a multiple quantum structure
and/or to a series of quantum dots. In these size induced quantum
wells, specific states are confined. One dimensional multiple
quantum well structures generated by compositional modulation of
nanowires were examined previously. For example, superlattices of
GaAs/GaP \cite{gudiksen}, InAs/InP \cite{bjork} and Si/SiGe
\cite{wu} nanowires were fabricated. Moreover, superlattices of Si
and Ge were investigated theoretically.\cite{nurten} In these
structures multiple quantum wells are formed because of the
different nature of materials that constitute the nanowire
superlattice. In the present work however, multiple quantum wells
are formed because of the quantum size effect, which is a
diversification of the electronic structure of the same material
with a change in its size.

Our results are obtained by performing first-principles plane wave
calculations within Density Functional Theory (DFT)\cite{kohn}
using ultrasoft pseudopotentials.\cite{vander,vasp} The exchange
correlation potential has been approximated by Generalized
Gradient Approximation using PW91 functional\cite{gga}. A
plane-wave basis set with kinetic energy cutoff of up to 250 eV
has been used. All atomic positions and lattice constants are
optimized. The convergence for energy is chosen as 10$^{-5}$ eV
between two steps, and the maximum force allowed on each atom is
0.02 eV/\AA. More details about calculations can be found in
Ref[\onlinecite{halfmetal}].

\begin{figure}
\includegraphics[width=8.4cm]{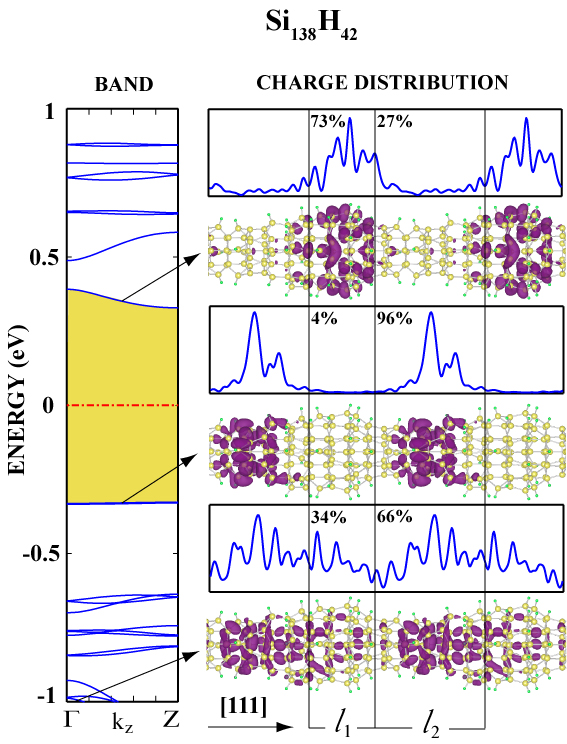}
\caption{Energy band structure of the superlattice,
Si$_{138}$H$_{42}$, formed from periodically repeated junctions
consisting of one unit cell of SiNW with $D_1$ $\sim$ 11 \AA~and
two unit cell of SiNW with $D_2$ $\sim$ 7 \AA~formed along [111]
direction. Here the diameter $D$ is defined as the largest
distance between two Si atoms in the same cross-sectional plane.
Isosurface charge densities of specific bands and their planarly
averaged distribution along the superlattice axis together with
their confinement in percentages are also shown. Large and small
balls indicate Si and H atoms. Zero of energy is set at the Fermi
level shown by dash-dotted line.} \label{sinwfig:1}
\end{figure}

\begin{figure}
\includegraphics[width=8.4cm]{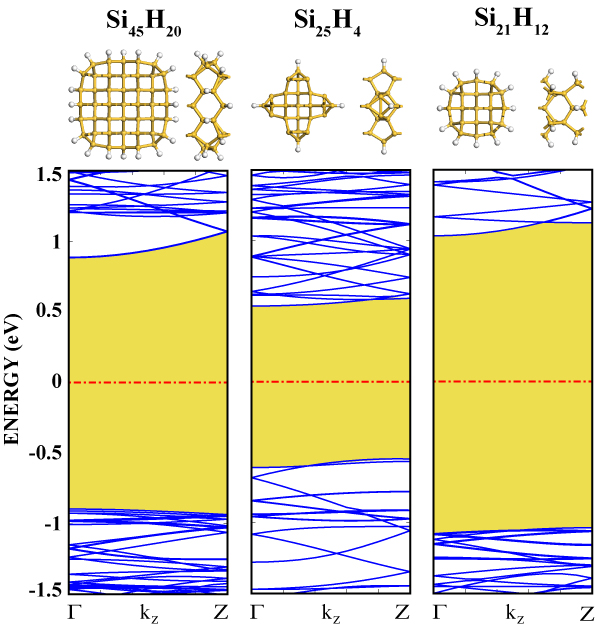}
\caption{Band structures of infinite and periodic
Si$_{45}$H$_{20}$, Si$_{25}$H$_{4}$ and Si$_{21}$H$_{12}$
nanowires, which constitute Si$_{157}$H$_{64}$ structure. The
front and side view of these relevant structures are shown on the
top of the band structure plots. Numerals as subscripts of Si and
H indicate the number of Si and H atoms in the unit cell. Energy
band gaps are shaded.} \label{sinwfig:2}
\end{figure}

\section{Size Modulated Silicon Nanowires}
Here Si nanowires (i.e. Si$_{N_{1}}$ with diameter $\sim D_1$ and
having $N_1$ Si atoms in the primitive unit cell) are first cut
from the ideal bulk crystal along desired direction. Then in every
alternating segment comprising $l_1$ unit cell the diameter $D_1$
is kept fixed, but in the adjacent segment comprising $l_2$ unit
cell (each having $N_2$ Si atoms) the diameter is reduced to
$D_2$. The latter part can be identified as the segment of
Si$_{N_2}$ nanowire. At the end, the segments of Si$_{N_1}$ and
Si$_{N_2}$ have made a smooth junction and hence formed an ideal
superlattice Si$_N$ so that its diameter is modulated in real
space. Note that $N\leq{l_1} {N_1}+{l_2}{N_2}$, because some
surface atoms attaching with a single bond to the surface were
removed at the beginning. Subsequently we relaxed the atomic
structure of this bare Si$_N$. Upon relaxation, the dangling bonds
on the surface are saturated by H atoms and Si$_N$H$_M$
superlattice is further relaxed for final atomic structure and the
lattice constant. The resulting superlattice can be described by a
Si rod with alternating diameters or a nanowire with alternating
wide and narrow parts. Here we consider two superlattice
structures, namely Si$_{138}$H$_{42}$ and Si$_{157}$H$_{64}$,
which are grown in [111] and [100] directions, respectively. For
the latter we investigate the surface defect and also boron, B and
phosphorus, P substitutional impurities.

\begin{figure*}
\includegraphics[width=12cm]{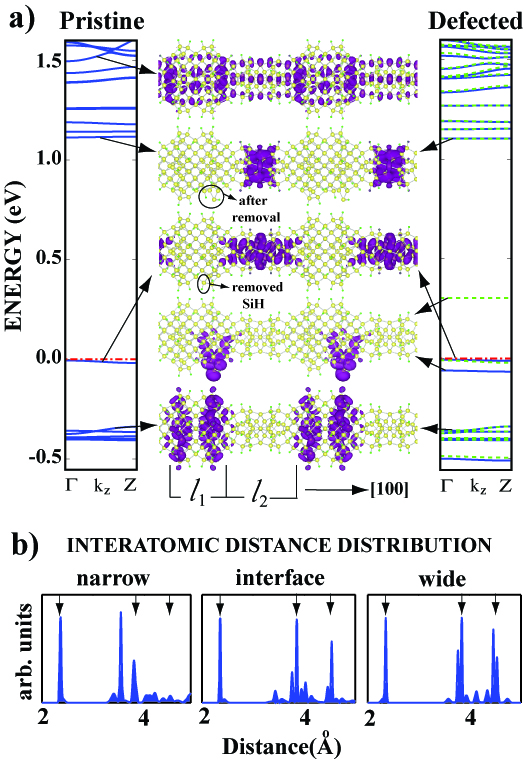}
\caption{(a) Electronic structure and isosurface charge densities
of selected states for pristine Si$_{157}$H$_{64}$ and defected
Si$_{156}$H$_{63}$ (with one Si-H on the surface removed)
nanowire superlattices. Arrows pointing at the same isosurface
charge density plot indicates that those states are not affected
by the formation of the defect and have nearly the same charge
density profile. Fermi level, represented by dashed-dotted (red)
line, was shifted to the valence band edge and set to zero. Solid
(blue) and dashed (green) lines in the right hand side box
represent the majority and minority spin bands, respectively. (b)
The distribution of interatomic distances of relaxed superlattice.
The arrows indicate the ideal bulk 1st, 2nd and 3rd nearest
neighbour distances.} \label{sinwfig:3}
\end{figure*}

\begin{figure}
\includegraphics[width=8.4cm]{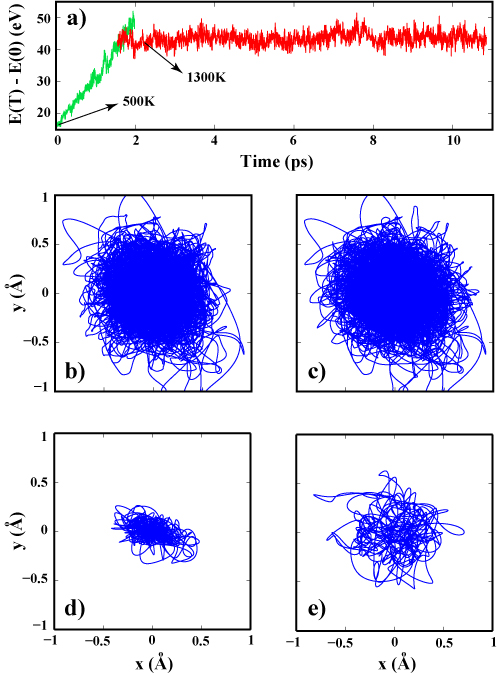}
\caption{(a) Energy per unitcell of Si$_{157}$H$_{64}$
superlattice as its temperature is increased from 500 K to 1400 K
(green/light) and kept constant at 1300 K (red/dark). Trajectories
of atoms at the surface of (b) wide and (c) narrow parts and at
the center of (d) wide and (e) narrow parts of the superlattice,
projected on a plane perpendicular to the wire axis, as the system
is kept at 1300 K. } \label{sinwfig:4}
\end{figure}

\subsection{Si$_{138}$H$_{42}$ Superlattice}

To form the superlattice, Si$_{138}$H$_{42}$, we took ${\it
l}_1=1$ unitcell of Si$_{68}$ and ${\it l}_2=2$ unitcell of
Si$_{38}$ cut in [111] direction of the bulk silicon crystal. The
combined structure had six Si atoms attached to the surface by
only one bond. These atoms were removed at the beginning and the
structure was optimized. After the relaxation, all Si atoms were
at least triply coordinated. The dangling bonds of triply
coordinated Si atoms were saturated with hydrogen atoms and the
structure was relaxed again. The band structure and isosurface
charge densities of the resulting superlattice is shown in
Fig.~\ref{sinwfig:1}. Propagating states, with dispersive bands
have charge densities distributed everywhere in the superlattice
rod. For example, a state near $\sim$ -1 eV is a propagating
state. The states of flat degenerate band at the top of the
valence band are confined in the narrow segment. The integral of
the planarly averaged charge,
$\int_{{\it{l}}_1}|\overline{\Psi({z})}|^{2} d{z}=0.96$ indicates
a rather strong confinement. The states of the lowest conduction
band with low dispersion are localized in the wide segment with
relatively weaker confinement. This situation implies the
staggered band alignment with confined hole states in the narrow
part and confined electrons in the wide part. Confinement may
increase with the size (length) of the confining segment since the
energy of the state relative to the barrier varies. For perfect
localization, i.e. $\sim$ 100 $\%$ confinement in ${\it l}_1$ or
${\it l}_2$, where coupling between confined states are hindered,
the segment ${\it l}_1$ or ${\it l}_2$ behaves as a quantum dot.
Confinement and band edge alignment in the present superlattice
are reminiscent of the 2D pseudomorphic or commensurate
semiconductor superlattices.\cite{Esaki}

\begin{figure*}
\includegraphics[width=15cm]{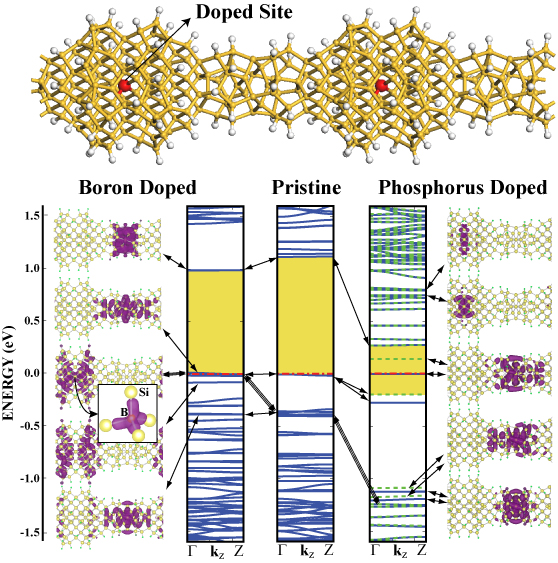}
\caption{Atomic and electronic structure and charge density
isosurfaces of Si$_{157}$H$_{64}$ substitutional doped by B and P
atoms (namely Si$_{156}$P(B)H$_{64}$). Top panel presents a view
of the superlattice structure, where the doped site is represented
by large red(dark) ball. Arrows between the bands of the undoped
(pristine) superlattice and the bands of the B and P doped
superlattices indicate the states, which preserved their character
even after doping. The charge density isosurfaces of specific
states are linked to the corresponding bands by arrows. Fermi
level, represented by dashed-dotted (red) line, was shifted to the
valence band edge and set to zero. Solid (blue) and dashed (green)
lines represent the majority and minority spin states,
respectively. Small panel at the left hand side presents the
charge density isosurface plot of B-Si bonds.} \label{sinwfig:5}
\end{figure*}
\subsection{Pristine Si$_{157}$H$_{64}$ Superlattice and Vacancy}

Here we presents the superlattice formation and confined states in
a different structure, namely Si$_{157}$H$_{64}$. To form this
structure we took ${\it l}_{1}=2$ unitcell of Si$_{45}$ and ${\it
l}_{2}=3$ unitcell of Si$_{25}$ cut in [100] direction of the bulk
silicon crystal. The combined structure had eight silicon atoms
that were attached to the surface by only one bond. We have
removed these atoms at the beginning and relaxed the resulting
structure. Then we have passivated the dangling bonds of the final
structure and relaxed the whole structure again. As a result, the
wide part of the nanowire superlattice had ${\it l}_{1}=2$
unitcell of Si$_{45}$H$_{20}$ with $D_1$ $\sim$ 14 $\AA$ and the
narrow part had ${\it l}_{2}=3$ unitcell of Si$_{25}$H$_{4}$ and
Si$_{21}$H$_{12}$ mixture with $D_2$ $\sim$ 7 $\AA$.
Figure~\ref{sinwfig:2} presents the band structure of the
corresponding infinite periodic nanowires which constitute the
Si$_{157}$H$_{64}$ superlattice structure. To compare with the
band structure of the superlattice (presented in
Fig.~\ref{sinwfig:3} as pristine), the band structure of these
nanowires were folded ${\it l}_{1}+{\it l}_{2}=5$ times. One can
see that the band gap of the superlattice is close to the lowest
band gap of its constituents. This is expected for the
superlattice structures having normal band lineup. The band gaps
of structures shown in Fig.~\ref{sinwfig:2} do not obey the
well-known trend, $E_{G}=C/D^{\alpha}+E_{G}[bulk Si]$, occurring
due to the quantum confinement effect\cite{confine} at small $D$. We 
ascribe this result to the surface reconstructions which differ 
in accordance with the procedure of atomic relaxation. Quantum confinement 
trend occurs when Si nanowires, cut from ideal bulk, are directly passivated 
with H atoms and subsequently relaxed. In the present study, however, the 
ideal structures are first relaxed, then passivated with H atoms and then 
relaxed again to obtain the final structure.\cite{halfmetal} Such Si 
nanowire structures with D$\sim$1 nm may not obey the quantum confinement 
trend. The decline of the gap values from the $1/D^{\alpha}$ behaviour 
has been also reported for other types of SiNWs and was attributed 
to surface effects.\cite{ponomareva}

Figure~\ref{sinwfig:3}(a) presents the band structure and the
projected charge density isosurface plots of pristine
Si$_{157}$H$_{64}$ and defected Si$_{156}$H$_{63}$ nanowire
superlattices. Flat minibands of Si$_{157}$H$_{64}$ structure near
the edge of conduction and valence bands are
distinguished.\cite{Esaki} In contrast to propagating states of
dispersive bands near 1.5 eV, the states of these flat mini bands
are confined in the narrow segments of the superlattice. This
situation also implies a normal band lineup. We are aware of the
fact that fabrication of nanowire superlattices having ideally
tapered as in our case is impossible. So we examined the effect of
an imperfection on the behavior of the system by removal of a Si-H
atom pair, which was normally attached to the surface. The
location of this defect and structural change after relaxation is
given in Figure~\ref{sinwfig:3}(a). The resulting structure has
odd number of electrons, so we made spin polarized calculation.
One can see that the valence and conduction band edges are not
affected by defect formation, while one empty and one filled state
localized around the defected area appear near the valence band
edge. The magnetic moment induced by the defect formation is 1
$\mu_B$. The vacancy states in bulk Si are located deep in the
band gap. In the superlattice structure, this vacancy state is
located near the valence band, since the defect is formed on the
surface at the interface between two segments having different
diameters.

In Figure~\ref{sinwfig:3}(b) the distribution of interatomic
distances shows significant reconstruction and deformation. In
particular, owing to increased surface/volume ratio the deviation
of the second and third nearest neighbour distances from the bulk
crystal is rather large at the narrow and interface regions. We
note that in both short periodicity superlattices discussed above
the electronic structures are significantly different from the
constituent SiNWs. As ${\it l}_1$ and/or ${\it l}_2$ increases,
the superlattice effects and confined states may become more
pronounced, and band-offset converges to a well-defined value.
When the lengths of ${\it l}_1$ and ${\it l}_{2} > \lambda_{B}$
(Broglie wavelength), the electronic properties in each segment
becomes close to those in constituent nanowires. In addition to
${\it l}_1$ and ${\it l}_2$, $D_1$ and $D_2$ and also the
direction of growth are also critical parameters, since the
electronic structure of the constituent SiNWs is strongly
dependent on these parameters. As $D_1$ and $D_2$ increases, the
surface effects decrease and the density of states becomes more
bulk like. The discontinuities of valence and conduction bands of
the superlattice is expected to diminish, since the band gap of
the constituent nanowires converge to the bulk Si value. However
for a significant $\Delta D=D_1-D_2$ the superlattice effects, in
particular confinement of specific states are expected to
continue. In contrast, smooth interfaces leading to hornlike
profiles can be used for focused electron emission.\cite{ciraci}
Recently, transport properties of size modulated SiNWs have been
shown to depend on the growth direction.\cite{nanoletter}

The stability of the Si$_{157}$H$_{64}$ structure was analyzed
using the finite temperature ab initio molecular dynamics
calculations with time steps of $2 \times 10^{-15}$ seconds.
Figure~\ref{sinwfig:4}(a) presents the total energy difference of
the system with respect to the ground state energy at 0 K as a
function of time. In the first calculation we have raised the
temperature of the system from 500 K to 1400 K in 2 ps. During
this calculation there was no major change in the structure and
the interface region was not destroyed by diffusion. In the second
calculation we took atomic positions and velocities of the system
from the first calculation when the system was at 1300 K and kept
the temperature constant for 9 ps. This calculation also resulted
in no major change in the structure of the superlattice.
Figure~\ref{sinwfig:4}(b) and (c) present the path of Si atoms
during the second calculation at the surfaces of wide and narrow
parts respectively. The range of the thermal fluctuations are
nearly the same and span a diameter of about 1 $\AA$.
Figure~\ref{sinwfig:4}(d) and (e) present the path of Si atoms at
the center of the wide and narrow parts respectively. As seen the
range of thermal fluctuations in the narrow part ($\sim 0.9 \AA$)
is larger than that in the wide part ($\sim 0.5 \AA$). It should
be noted that, the temporal trajectories of 9 picoseconds in the
present ab initio molecular dynamics calculations are rather long
as compared to ones usually carried out in the stability tests,
but may not be long enough to accumulate necessary statistics.
However, the temporal trajectories at temperature as high as 1300
K in the present study are long enough to eliminate the existence
of any atomic configuration in a weak local minimum of the
Born-Oppenheimer surface, which is prone to structural
instability. Therefore, based on the accurate structure
optimization and temporal trajectories as long as 9 picoseconds
carried out at 1300 K, we believe that superlattice structure
under study is stable at least at a temperature higher than the
usual operation temperature ($\sim$425 K) of a device.

\subsection{B and P doping of Si$_{157}$H$_{64}$ Superlattice}

Next we examine the effect of doping of the Si$_{157}$H$_{64}$
superlattice by substitutional B and P impurities. Both structures
have odd number of electrons and hence spin polarized calculations
have been carried out. Nevertheless, B doped structure resulted in
a nonmagnetic state with half filled band edge. P doped structure
resulted in a magnetic state with a magnetic moment of 1$\mu_B$.
In Fig.~\ref{sinwfig:5} the electronic structure of pristine and
doped superlattices are shown to follow the effect of the doping
on the states at the band edges. Owing to the small diameter of
the superlattice, the band structure is affected by the impurity.
Since the band gap of the wide part is larger than that of the
narrow part, the state associated with the impurity, which occurs
normally either near the valence or conduction band, falls in the
bands of the narrow part.

It is known that substitutional B doping of bulk silicon crystal
results in an acceptor state $\sim$45 meV above the valence band
edge.\cite{handbook} In the substitutional doping of the
superlattice by B similar situation occurs and the acceptor state
is located above the valence band edge corresponding to the wide
part, but falls in the valence band of the narrow part. The filled
first and the lower lying three valence bands at about $\sim$-0.4
eV undergo changes: The filled first valance band having states
confined in the narrow part becomes half-filled. Three lower lying
valence bands of the pristine superlattice are raised towards the
edge of the valence band edge. The states of these bands mix with
the tetrahedrally coordinated $sp^3$ hybridized orbitals of B and
form a triply degenerate state, just below the half filled valence
band edge. Charge density isosurfaces of these states around boron
atom presented in Fig.~\ref{sinwfig:5} clearly indicates $sp^3$
hybridization in tetrahedral directions.  As a result, the "hole
like states" becomes confined in the narrow part. As for the
conduction band is slightly shifted down, but its states preserve
their character.

In P doping of the superlattice at the center of the wide part,
the impurity state localized at P atom occurs in the conduction
band of the superlattice. This state, in fact, the donor state
occurring $\sim$ 200 meV below the conduction band corresponding
to the wide part. This energy is in good agreement with the energy
calculated recently for the ionization energy of P donor states in
the Si nanowires.\cite{doping} Below this donor state, all
conduction band states are confined either in the narrow part or
at the interface, but not in the wide part. Energy shifts of the
band edge states are more pronounced in structure having
substitutional P doping. Conduction band edge is shifted down by
0.75 eV, but the charge density profile remains the same. One
spin-up and one spin-down bands split by $\sim$0.2 eV are formed
near the conduction band edge with charge densities localized in
the narrow part. The triply degenerate state in the valance band
near the edge are shifted down. That states localized around
substitutional P impurity appear in the conduction band continua,
while a filled spin up band states are localized in the defect free
narrow part, indicate the possibility of modulation doping. In
closing this section we note that the electronic structure
associated with the doping of the small diameter superlattices is
complex and has aspects different from that in the doping of bulk
Si. The small size and dimensionality of the doped superlattices
hinders the application of the effective mass approximation
leading to hydrogenic impurity model. For the same reason, the relative 
positions of band edges can be affected by the presense of a substitutional 
impurity.

\section{Conclusion}
In conclusion, we considered two specific superlattices grown in
different crystallographic directions as a proof of concept. Each
superlattice can be viewed to form the periodically repeating
junctions of two hydrogen saturated silicon nanowires of different
diameters. Since the band gaps of the constituent Si nanowires are
different, the band gaps of these superlattices are modulated in
real space. As a result, one dimensional multiple quantum well
structures with states confined in different (wide or narrow)
parts are formed. At strong confinement or localization, the superlattice structures can be viewed as a series of quantum dots. We believe that the bad gap modulation and resulting quantum well structure
and confinement can occur as long as the band gap differences are
maintained between adjacent region at both sides of the junction.
Furthermore, we showed that the electronic structure resulting
from the substitutional doping of the superlattice interesting and
give rise to modulation doping. We finally note that, the band gaps of  SiNWs  at both sides of junction forming the superlattice structure are underestimated by DFT/GGA calculations. Since the systems under study are too large, these bands cannot be improved by GW calculations. Nevertheless, our work is just a proof of concept. According to our results, if the band gaps are different in different parts of the heterostructure, confined states can occur and modulation doping of these quantum structures can be achieved. Details of energy band structure and location of impurity states in the gap should only be taken qualitatively.

Various types of electronic devices, such as resonant 
tunnelling double barriers generated from these superlattices, 
can be arranged on a single rod, where
they can be connected by the metallic segments of unsaturated Si
nanowires. The character of these devices can be engineered by
varying their growth direction and structural parameters, such as
${\it l}_1$, ${\it l}_2$, $D_1$, $D_2$. Even if the diameters of
superlattices we treated in this study are small ($\sim$ 1 nm) and
hence in the range where band gaps vary significantly with
diameter, they are still in the range of those fabricated
recently.\cite{Ma}

\section{Acknowledgement}
Part of the computations have been carried out by using UYBHM at
Istanbul Technical University through a grant (2-024-2007). This
work is partially supported by T{\"U}BA, Academy of Science of
Turkey.

\end{document}